\documentclass[aps,pra,twocolumn,longbibliography,superscriptaddress,10pt]{revtex4-1}   
\usepackage{graphicx}             
\usepackage{hyperref}
\usepackage{amsmath}

\begin{document}
\title{Transport of spatial quantum correlations through an optical waveguide}   
\author{J. Hordell} 
\affiliation{Midlands Ultracold Atom Research Centre, School of Physics and Astronomy, University of Birmingham, Edgbaston, Birmingham B15 2TT, UK}
\author{D. Benedicto-Orenes}
\affiliation{Midlands Ultracold Atom Research Centre, School of Physics and Astronomy, University of Birmingham, Edgbaston, Birmingham B15 2TT, UK} 
\author{P. G. Petrov}
\email{p.g.petrov@bham.ac.uk}
\affiliation{Midlands Ultracold Atom Research Centre, School of Physics and Astronomy, University of Birmingham, Edgbaston, Birmingham B15 2TT, UK}
\author{A. U. Kowalczyk}
\affiliation{Midlands Ultracold Atom Research Centre, School of Physics and Astronomy, University of Birmingham, Edgbaston, Birmingham B15 2TT, UK} 
\author{G. Barontini}
\affiliation{Midlands Ultracold Atom Research Centre, School of Physics and Astronomy, University of Birmingham, Edgbaston, Birmingham B15 2TT, UK} 
\author{ V. Boyer}
\email{v.boyer@bham.ac.uk}
\affiliation{Midlands Ultracold Atom Research Centre, School of Physics and Astronomy, University of Birmingham, Edgbaston, Birmingham B15 2TT, UK}

\begin{abstract}
The ubiquity of optical communications is due in large part to the advent of the optical fibre, which allows for flexible and efficient routing of light-encoded information. Used as serial channels, single fibres have also been shown to be effective to transport quantum information, for instance in commercial quantum key distribution systems \cite{peev}. As fibre technology progresses to support the transmission of full images, e.g. in endoscopic devices, the question arises whether this technology is also suitable for the parallel transport of spatial quantum information, such as quantum images \cite{boyer_prl_2008}. Here we demonstrate the transport of quantum intensity correlations through a conduit made of the ordered packing of thousands of fibres, in a way which preserves localised intensity-difference squeezing. Maintaining the spatial character of quantum information opens the way to the use of guided-light technology in the emergent field of quantum imaging.
\end{abstract}

\maketitle

Optical fibre bundles are readily used to transport information in the form of classical images, for example in endoscopy devices \cite{subramanian}. In these applications, the information is contained in the local amplitude of the light field incident on the input face of the bundle. The ordered packing of the fibres in the bundle guaranties the conservation of the light intensity distribution across the beam profile, therefore the image is preserved during the propagation and reappears on the output face of the bundle. The question remains whether one can extend this property to the quantum domain. This will be equivalent to ensuring that the quantum correlations associated with the intensity fluctuations of an arbitrary spatial feature, that is to say an input transverse mode, are reproduced on the output of the guide in the exact same mode.

Previous experimental research in the guiding of the spatial quantum states of
light has been essentially limited to the single-photon subspace. For instance, the guided transport of photons spatially entangled over two modes has been demonstrated in a Kagome fibre~\cite{loffer}. Beyond a small number of modes,
only the spatially-resolved detection of photons has been demonstrated,
typically by combining a fibre bundle to an ensemble of single-photon detectors.
In this manner, the temporal nonclassical statistics of spatially resolved
photons emitted by scattered emitters was observed over a bundle of 15
fibres~\cite{israel}. Such spatially resolved guiding of single photons, or
photons belonging to correlated pairs, could be useful to channel
illumination or signal in low-light-level quantum imaging
applications~\cite{brida}. The interest in transporting quantum states of light for
imaging however, goes beyond these low fluxes of photons. Macroscopic states of
light are better suited to the imaging of hard-to-see objects, e.g. very weakly
absorbing objects, because at the quantum noise limit (QNL) a larger probing intensity
provides a better signal-to-noise ratio. In this context, bright two-mode
squeezed states of light, containing arbitrary mean numbers of photons,
have been shown to display local intensity correlations below the shot noise
limit both in the frequency~\cite{boyer_prl_2008} and time domains~\cite{kumar},
and are therefore good candidates for practical quantum imaging. This is
particularly true for those applications where the illumination intensity cannot be arbitrarily
increased due to the existence of a damage threshold~\cite{taylor}, and where
squeezing the quantum fluctuations of light are the only solution to increase
the signal to noise ratio.

Here we address the issue of the transport of a bright illumination with a fibre bundle while preserving local intensity fluctuations at the quantum level. 
Starting from a two-mode squeezed state
\cite{heidmann}, that is to say a pair of continuous-variable (CV) entangled
light beams which display intensity and phase quantum correlations of their
random spatial fluctuations, we transport one of the beams through the bundle
and show that the spatial intensity correlations with the other beam are
preserved, as evidenced by the presence of local intensity-difference squeezing
after the transport.

The two-mode squeezed state is created using a phase-insensitive optical amplifier. Our amplifier is based on four-wave-mixing (4WM) in a hot $^{85}\mathrm{Rb}$ vapour in a double-lambda configuration~\cite{mccormick}, as described in the Methods section. The process is efficient enough to realise a nonlinear single-pass gain of a few units without the need of an enhancement cavity. This travelling-wave amplifier, because it has a gain region of finite length, does not enforce a strict phase-matching condition and the resulting two-mode squeezed state is intrinsically spatially multimode. This means that the twin beams, probe and conjugate, are entangled across multiple spatial modes, or regions~\cite{boyer_prl_2008}. In particular, matching pairs of locales in the cross-sections of the beams are correlated. In the plane of the cell, referred to as the near field, these regions are mapped onto each other because probe and conjugate correlated fluctuations are born in the same location. In the far field these regions are symmetrically placed about the pump propagation axis due to transverse momentum conservation. 
Diffraction associated with the propagation over the finite length of the cell leads to a minimum transverse size over which correlations can be observed, the coherence area~\cite{brambilla}, with an associate coherence length \footnote{This a transverse coherence length. and the only one we will refer to in the rest of the paper}. As the coherence area is substantially smaller than the beams size, a large number of modes, typically a few tens, are independently correlated. This unique feature makes the twin beams a useful resource for imaging with sensitivity \cite{lopaeva} or resolution \cite{taylor_prx} beyond the shot-noise limit, and efficiently guiding of the beams would make these applications more practical.

\begin{figure}[t!]
  \centering
	  \includegraphics[width = 8cm]{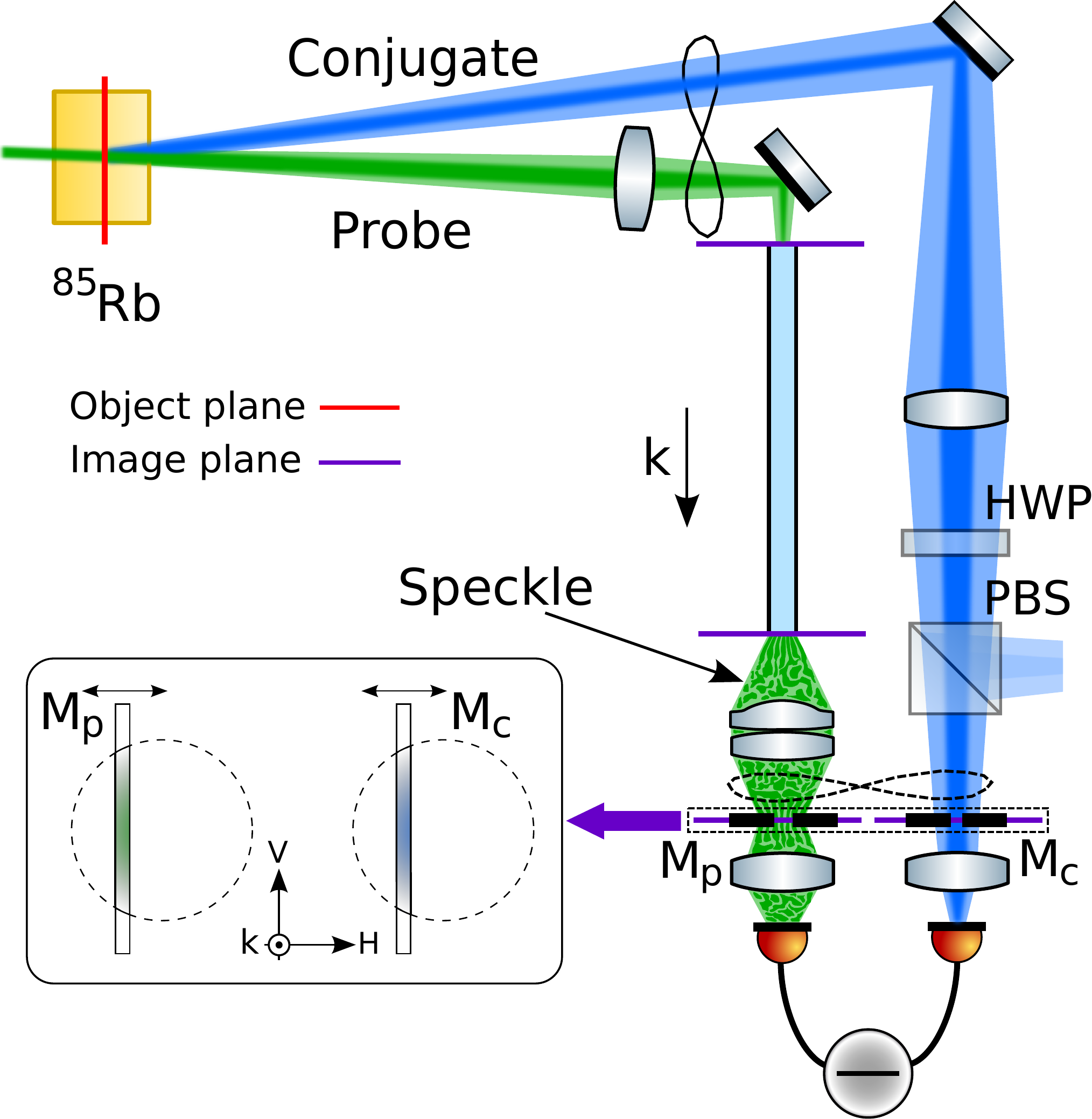}
	\caption{
      \textbf{Experimental setup.} The conjugate and probe beams are generated via 4WM in the $^{85}$Rb cell. 		  The probe light in the object plane, i.e. the plane at the position
      of the cell, where the correlations are created, is imaged onto the input
      face of the conduit. The probe light emerging from the output face of the
      conduit is imaged onto the probe slit on the measurement plane
      ($\mathrm{M_p}$) using a high numerical aperture optical system. The
      conjugate is directly imaged onto the conjugate slit on the measurement
      plane ($\mathrm{M_c}$). The conjugate slit is translated in horizontal (H)
      and vertical (V) directions perpendicular to the direction of propagation
      $\mathrm{\bf k}$, for every position of the probe slit as shown in the
      inset. The polarising beamsplitter (PBS) and the half-wave plate (HWP)in the path of the conjugate provide a tunable loss for
      the channel to balance the beams intensities in the case where the probe
      is sent through the conduit. The lightly coloured paths depict the geometric
      propagation of the quantum fluctuations whereas the densely coloured paths
      depict the Gaussian propagation of the bright probe and conjugate beams.
      The pump beam is 
      not shown for clarity. The figures of eight indicate CV entanglement (solid 
      line) and quantum intensity correlations (dashed line).}
		\label{fig:ex-setup}
\end{figure}

We chose to focus on intensity correlations in the near field. To this effect,
we separate the twin beams and we image their nonlinear interaction region onto
individual ``measurement'' planes, as shown in Fig.\ref{fig:ex-setup}. In these
planes, identical locations on the twin beams display quantum intensity
correlations which we aim to preserve after transport in a fibre bundle.

\begin{figure}[t]
  \centering
	  \includegraphics[width = 8cm]{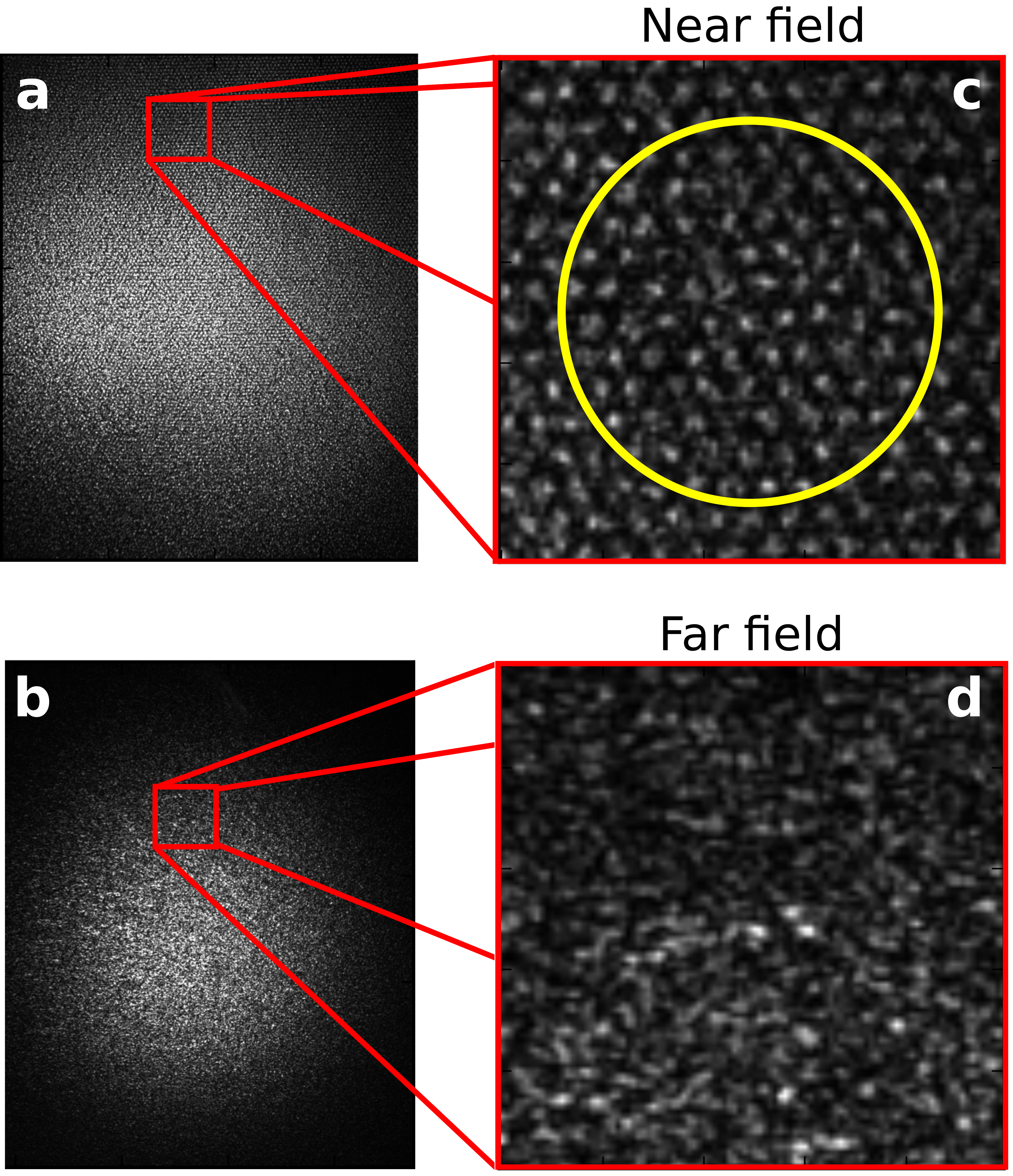}\\
		\caption{
      \textbf{Images of the probe beam after the conduit.} \textbf{a,b} Images of the near-
      and far-field beam patterns after the conduit, showing
      the Gaussian intensity envelope. \textbf{c,d} Selected magnified
      regions across the profiles in \textbf{a} and \textbf{b}, respectively. 
      In \textbf{c}, in the
      near-field, the characteristic pattern coming from the hexagonal lattice
      of constituent fibres can be clearly seen at the output face of the
      conduit. The lattice spacing is about $\mathrm{12~\mu m}$. Since the
      fibres are multimode, the emerging mode shape is random across the fibre
      lattice. The enclosed circular area represents the estimated coherence
      area of the process. In \textbf{d}, in the far field the fibre lattice pattern is
      transformed into a speckle pattern as the light from every fibre
      acquires a random phase.}
		\label{fig:beams}
\end{figure}

The fibre bundle itself is a rigid conduit, designed to transmit (coherently) optical images with a resolution given by the fibre size. It is 15$\mathrm{cm}$ long and contains $5\times10^{4}$ fibres with diameter of 12$\mu\mathrm{m}$ tightly packed into a equilateral triangular lattice. Fig.~\ref{fig:beams}(a) shows an image of the output face of the conduit injected with a Gaussian beam of diameter smaller than the conduit diameter, and reveals the packing structure. The incomplete filling leads to an overall transmission of around 30\%, primarily determined by the ratio of the fibre core total area to the conduit input face area, but also by partial reflection on the faces of the conduit. The large core size of the fibres makes them multimode for near-infrared optical wavelengths, as evidenced by the irregularity in the shapes of the light beams coming out of the individual fibres.

The discretisation introduced by the finite fibre size sets a natural
length scale for the smaller spatial details that can be transmitted by the
conduit. At the quantum level, one can expect that only fluctuations of
spatial modes substantially larger than the inter-fibre spacing will be
accurately transmitted across the guide. We therefore image the beam on the input face of the conduit in such a way that the fibre size is much smaller than
the estimated coherence area, as
shown in Fig.\ref{fig:beams}(a). This means that all the scales of correlations present in the system should be successfully transmitted. 

The loss of probe light due to the finite transmission of the conduit induces a
mismatch between the probe and conjugate fluctuations at the balanced
photodetector, which can be corrected by introducing an adequate level of
attenuation on the conjugate beam, as shown in Fig.~\ref{fig:ex-setup}. Since
the effect of loss on a beam of light is to mix vacuum noise to the fluctuations
of the beam, the measured squeezing after loss is less than the squeezing before
loss. Therefore the initial degree of squeezing (3.0 dB) and the level of
attenuation of the conjugate (25\%) are chosen to maximise the level of
squeezing seen after the conduit. The details of the procedure are given in the Supplementary Information.

Although the fibre bundle conserves the coarse distribution of intensity between
the input and the output, modal
dispersion, i.e.\ group velocity dispersion for different spatial modes within
each fibre~\cite{yariv} as well as differential dispersion between the fibres,
leads to the scrambling of the phase of the output field on a length scale of the
order of the size of the fibre cores. As a consequence of this wavefront distortion,
the output beam is diffracted following a large numerical aperture (0.55) 
and produces in the far field the speckle pattern shown in
Fig.\ref{fig:beams}(b). Since the spatial correspondence of the intensity
between the input and output faces does not extend to regions away from the
faces, faithful transmission of the intensity fluctuations puts
stringent requirements on the accuracy of the imaging of the plane of interest
onto the input face of the conduit. Similarly, the output plane has to be 
accurately imaged onto the measurement plane. Because of the strong diffraction, 
the collection optics requires a high numerical aperture (0.5) and has to be accurately placed due to the resulting very short depth of field.

In practice, we test the spatial correlations by selecting small regions on the
measurement planes with two movable slits, one on the probe beam and one on the
conjugate beam (Fig.~\ref{fig:ex-setup}). The partial powers transmitted through
the slits are detected by the separate photodiodes of a balanced photodetector
which forms the subtraction of the photocurrents. The noise of the photocurrent
difference is measured with a spectrum analyser, and recording the
intensity-difference noise as a function of the position of the slits allows us
to map the spatial intensity correlations between probe and conjugate. Selecting
correlated regions leads to reduced noise below the QNL, whereas mismatched
positions produce excess noise associated with phase-insensitive optical amplification. We perform the experiment twice. Firstly, we characterise the spatial
correlations generated by the 4WM alone. Secondly, we measure the same spatial
correlations after sending the probe beam through the conduit, as shown in
Fig.\ref{fig:ex-setup}.

\begin{figure*}[t!]
	\centering
		\includegraphics[width=16cm]{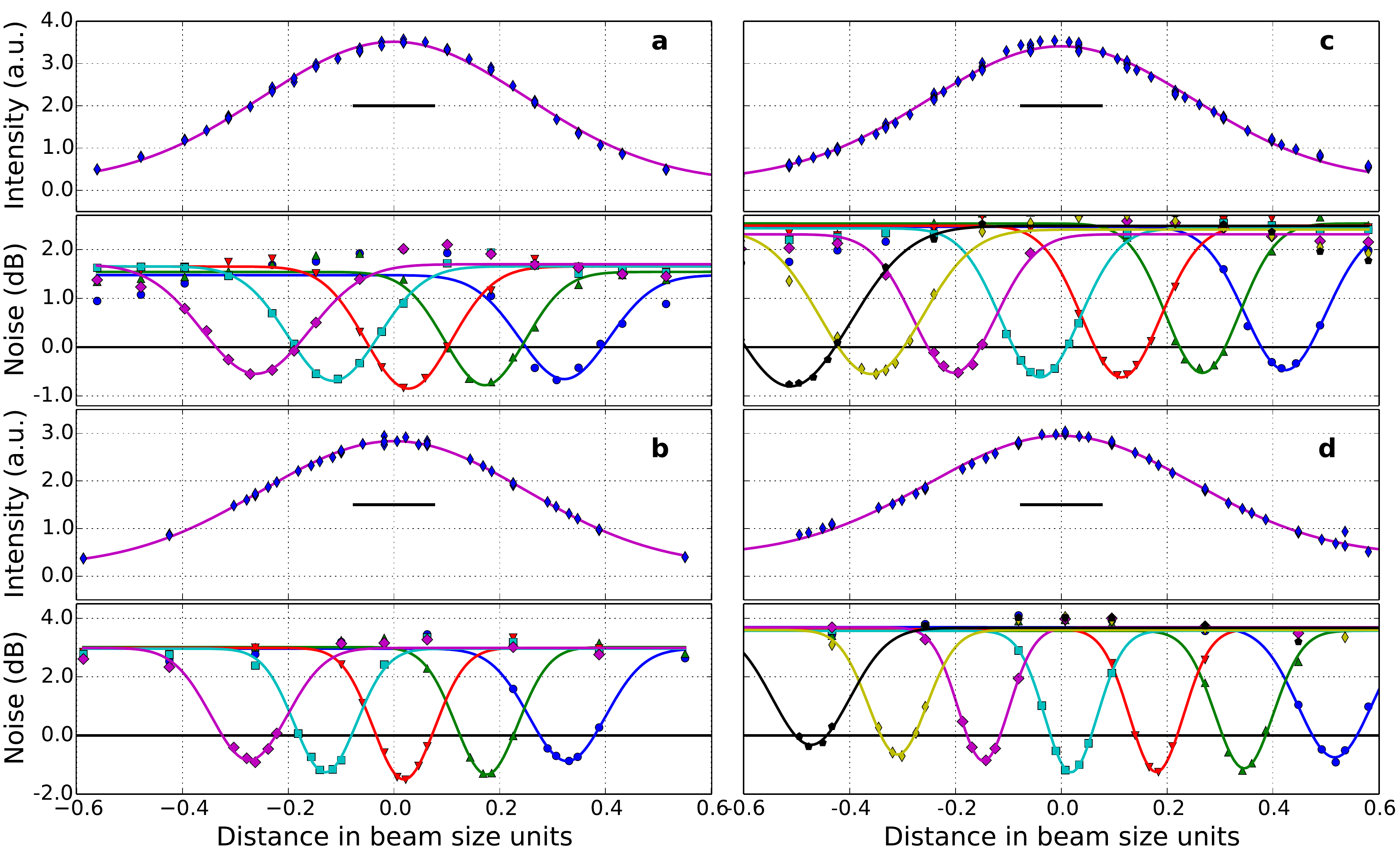}
    \caption{\textbf{Scanning noise data relative to shot noise.} The measurement
      procedure is done by scanning the slit position on the conjugate beam for
      fixed positions of the slit on the probe beam. The result is a single dip
      in the detected noise for each scan. The scanning is done in both
      horizontal \textbf{a,b} and vertical \textbf{c,d} directions for the cases where the
      conduit is inserted in the probe beam path in \textbf{a,c}, and when it is not
      present \textbf{b,d}. The top graph in each figure is the fitted intensity
      profile of the conjugate beam. The noise dips are fitted using Gaussians
      and the black solid line at 0 dB represents the shot noise. The bar in
      each beam profile graph denotes the slit size.
      The excess amplification noise, observed when the slits positions do not
      match, is larger at the centre of the gain medium, where the Gaussian pump
      beam profile peaks and the gain is the largest. The intensity profile of the 
      conjugate beam is given in arbitrary units (a.u.).}
	\label{fig:scanning_data}
\end{figure*}

The slits are set up on pair of translation stages and their width is set to be
of the order of the coherence length. While it is possible to accurately
evaluate the coherence length \cite{holtfrerich}, we adopt here a more
straightforward methodology which consists in selecting widths of the slits which
cause a significant reduction of measured squeezing. Reducing the slits sizes
increases the share of the spatial frequency spectra of the transmission
functions of the slits which lie outside the spatial bandwidth of the 4WM
process and corresponds to transmitted modes in a vacuum state. If this
fraction of the spectrum dominates, the measured noise will tend 
to the QNL. Note
that the spatial spectra associated with the gate-shaped transmission function
of the slits always contain high spatial frequencies that lie outside the 4WM
spatial bandwidth, which is given by the coherence length. 
It is therefore expected that selecting correlated regions
with slits will always generate reduced measured squeezing. Experimentally the
slits are set to $\approx$15~\% of the beams widths and provide an overestimate of the
coherence length \cite{embrey}, theoretically estimated to be around 120~$\mu$m which is about 8~\% of the beams size.
After the width is selected, the slits are scanned across the span of the beams,
scanning over the conjugate for each region across the probe.
Fig.\ref{fig:scanning_data} shows the noise recorded by the balanced detector for
each pair of positions of the slits. A reduction in the intensity-difference
noise below the QNL over a background of excess noise is observed when the slits
positions match, that is to say they are identical inside the probe and conjugate
beams.

Comparing the data with and without the bundle, it is apparent that, apart from
a reduction of the overall level of squeezing due to the finite transmission of
the device, the conduit preserves the spatial properties of the quantum state of
the light, here the intensity quantum fluctuations. As expected, the spatial
resolution of the conduit is better than the coherence length of the beam profile. Indeed the fibre
size is 10 times smaller than the coherence length of the transmitted quantum
field, which, as estimated by the width of the narrowest trough in
Fig.\ref{fig:scanning_data}(d) is of the order of
$100~\mu\mathrm{m}$. The exact shape of the
noise dips is the result of a combination of the slits transfer function
and the coherence profile. The width of the dips can therefore be used to
estimate trends of the relative size of the coherence area with
respect to the beam sizes. For a fixed slit width, a wider noise dip implies a
larger coherence length. In order to gauge the effect of the conduit on
the spatial frequency bandwidth of the correlations, we consider the width of the squeezing troughs, $\kappa$, relative to the Gaussian beam diameter.
In the case of transmission through the conduit, $\kappa$ is found to be 0.32
and 0.29, for horizontal and vertical scans respectively. In the case of free
space propagation, $\kappa$ is 0.27 and 0.20, respectively.

The comparison between the values of $\kappa$ with and without the conduit
shows that the spatial bandwidth of the spatial correlations is largely
preserved during transport. The small loss of spatial resolution can be
attributed to a combination of imperfections in the imaging system and conduit
behaviour. Firstly, the very small depth of field of the high-numerical aperture
imaging system, which is required to efficiently collect the light after the conduit, could
result in an approximate imaging of the output face of the conduit on the slit. This works in
combination with the spherical aberrations of the optical system. Secondly, the
conduit itself has a small but not negligible cross-talk between
fibres~\cite{perperidis}, which leads to the smearing of the correlations
over adjacent correlated regions, resulting in larger width of the troughs.

The two-mode squeezed state generated by the 4WM process nominally displays
phase-sum squeezing which, together with the intensity-difference squeezing, results in continuous-variable entanglement~\cite{boyer,reid}. Phase
scrambling by the conduit prevents us from directly measuring phase
correlations, however phase randomisation due to modal dispersion is to be
understood with respect to an external reference. Indeed, phase information can
still be transmitted as long as a bright coherent carrier is also transmitted so
as to serve as a local phase reference, as it is done in this experiment by
using a bright beam. The beam emerging from the conduit should display phase 
fluctuations locally correlated with the phase fluctuations of the other twin beam. 
Measuring such correlations would require a local oscillator
with a wavefront which matches the wavefront of the light on the output face of the
conduit. Phase control of the local oscillator with a spatial resolution better
than the fibre array resolution would be a formidable yet feasible task.

Another solution to the issue of wavefront distortion is direct control of modal
dispersion, so that the transmitted beam emerges from the conduit with a flat
wavefront. Simply using single-mode fibres as the constituents of the bundle can
avoid modal dispersion within individual fibres, but controlling or correcting for
the differential dispersion between the fibres is much harder.
Nonetheless there has been progress in real time compensation of modal
dispersion to perform classical imaging through single multimode guides or
fibres \cite{porat,fertman,cizmar}. Improving substantially on the spatial
resolution of these techniques may lead to the possibility of guiding and
preserving the full quantum state of the electromagnetic field, thereby
extending the work of Ref.~\cite{loffer} to larger numbers of spatial modes and
arbitrary photon numbers.

In conclusion, we have demonstrated that spatial quantum intensity correlations,
in the form of localised intensity-difference squeezing, can be transported
through a fibre bundle. Our work provides initial step towards the realisation 
of practical quantum enhancement in imaging applications where the shot noise is a limitation and where light guiding provides additional convenience. Our experiment uses a standard commercially available conduit, however improved performance could be obtained with a purpose-build device made of single mode fibres for a reduced distortion of the wavefront, more fibres for improved spatial resolution, and optimised packing for increased transmission. Additionally, realising active phase control would extend the method presented here to the effective guiding of massively parallel entanglement, with possible use in parallel quantum communication links. 

\section*{Methods}

\textbf{Four-wave-mixing}. The quantum correlations are generated using a
non-degenerate 4WM process in a hot $^{85}\mathrm{Rb}$ atomic vapour in a
double-lambda configuration~\cite{mccormick}. A strong 795~nm pump beam (700~mW) and a weak seed beam (130~$\mu\mathrm{W}$) intersect at a small
angle of $\sim6\;\mathrm{mrad}$ into a 12 mm-long vapour cell heated to
$\gtrsim100\mathrm{^{\circ}C}$. The probe beam is derived from the pump beam via
an AOM in a double-pass configuration driven at a frequency of 1.5~GHz, half the ground state
hyperfine splitting. The pump and the probe are resonant with a two-photon Raman
transition between the hyperfine ground states, with a detuning of $\sim 800$
MHz to the blue of the $F=2\rightarrow F'$ and $F=3\rightarrow F'$ transitions,
respectively.

The 4WM process corresponds to the transfer of an atom from the $F = 3$
hyperfine ground state to the $F = 2$ hyperfine ground state and back to the
initial state. In the process the atom consumes two pump photons and emits one
probe photon as well as one conjugate photon at a frequency $\sim 6$ GHz higher
than the probe frequency and located symmetrically with respect to the pump beam
axis \cite{mccormick}. The hot atomic vapour acts as a phase-insensitive
amplifier for the probe with a single-pass gain of 1.5-2. This is enough to amplify the probe and generate a conjugate
beam in a travelling-wave configuration, that is to say without the use of a
cavity. The bandwidth of the nonlinear process is about 20~MHz.

\textbf{Optical imaging}. To reveal the intensity spatial correlations produced
locally in the gain medium, the median plane in the vapour cell, the near field,
must be imaged on the selection slits. When the conduit is introduced in the
probe path, the near field must in addition be imaged on the input facet of the
conduit. Direct imaging of the near field on the conduit or the slits is done
with a single long-focal-length lens of 250~mm. Imaging of the exit facet of the
conduit on the probe selection slit is done with high-numerical-aperture doublet
made of 8~mm focal length aspheric lens and a 16~mm focal length lens.

\textbf{Noise detection}. The light going through the slits is detected by a
balanced photodetector with quantum efficiency of 95\%. For practical reasons,
the gain is identical for both photodiodes. The trans-amplified differential
photocurrent is analysed with a spectrum analyser operating at an analysing
frequency of 2~MHz, a resolution bandwidth of 30kHz, and a video bandwidth of
30Hz.

\section*{Acknowledgements}
The authors acknowledge support from the Engineering and Physical Sciences Research Council, Grants No. EP/I001743/1 and No. EP/M013294/1. J. H. was supported by the Defence Science and Technology Laboratory research PhD program via contract No. DSTL-1000092268.

\section*{Author Contributions}
J.H., D.B. and A.K. constructed the apparatus, J.H. took the data, P.G.P. analysed the data and wrote the article with the help of V.B. The principal investigators V.B. and G.B. designed the experiment and provided guidance. All authors commented on the manuscript, discussed its structure, data analysis and interpretation.

\section*{Additional Information}
Supplementary information is provided. The authors declare no competing financial interests. Reprints and permission information is available online at www.nature.com/reprints. Correspondence and requests for materials should be addressed to P.G.P. and V.B.

		
\end{document}